\documentclass[prl,twocolumn,aps,amsmath,amssymb,nofootinbib,preprintnumbers]
{revtex4}

\usepackage{graphicx}
\usepackage{dcolumn}
\usepackage{bm}
\usepackage{amsmath}
\usepackage{amsfonts}

\def\ls{\mathrel{\lower4pt\vbox{\lineskip=0pt\baselineskip=0pt
           \hbox{$<$}\hbox{$\sim$}}}}
\def\gs{\mathrel{\lower4pt\vbox{\lineskip=0pt\baselineskip=0pt
           \hbox{$>$}\hbox{$\sim$}}}}
\def\drawbox#1#2{\hrule height#2pt

\hbox{\vrule width#2pt height#1pt \kern#1pt
              \vrule width#2pt}
              \hrule height#2pt}

\def\Asym#1#2{\vcenter{\vbox{\drawbox{#1}{#2}
              \kern-#2pt       
              \drawbox{#1}{#2}}}}

\newcommand{\beq}{\begin{equation}}
\newcommand{\eeq}{\end{equation}}
\newcommand{\bk}{{\bf k}}
\newcommand{\bq}{{\bf q}}
\newcommand{\bp}{{\bf p}}
\newcommand{\bx}{{\bf x}}

\newcommand{\TT}{{\rm TT}}

\newcommand{\nn}{\nonumber}
\newcommand{\GW}{{_{\rm GW}}}
\newcommand{\lb}{\left\lbrace}
\newcommand{\rb}{\right\rbrace}
\newcommand{\ab}{{\alpha\beta}}

\begin{document}

\title{Stochastic Background of Gravitational Waves from Fermions
}

\author{Kari Enqvist, Daniel G. Figueroa and Tuukka Meriniemi}

\affiliation{
Physics Department, University of Helsinki and Helsinki Institute of Physics,
P.O. Box 64, FI-00014, Helsinki, Finland.}

\date{March 22, 2012}

\begin{abstract}
Preheating and other particle production phenomena in the early Universe can give rise to high-energy out-of-equilibrium fermions with an anisotropic stress. We develop a formalism to calculate the spectrum of gravitational waves due to fermions, and apply it to a variety of scenarios after inflation. We pay particular attention to regularization issues. We show that fermion production sources a stochastic background of gravitational waves with a significant amplitude, but we find that typical frequencies of this new background are not within the presently accessible direct detection range. However, small-coupling scenarios might still produce a signal observable by planned detectors, and thus open a new window into the physics of the very early Universe.
\end{abstract}
\preprint{HIP-2012-09/TH}
\maketitle


{\bf Introduction.} The existence of gravitational waves (GW) is arguably one of the most important predictions of the general theory of relativity that still remains unverified. The measured decay of the orbital period of compact binaries~\cite{HulseTaylor} has nevertheless provided a strong, though indirect, evidence for GW. On theoretical grounds, we expect that the universe today should be permeated by a variety of GW backgrounds of diverse origin. For instance, GW are expected from astrophysical sources, like the collapse of supernovas or the coalescence of compact binaries. They are also expected from high-energy phenomena in the early Universe, see~\cite{Maggiore} for a review. Many plans for direct detection experiments exist, such as the VIRGO interferometer, the Laser Interferometer Gravitational Wave Observatory (LIGO), the European Laser Interferometer Space Antenna (eLISA), the Big Bang Observer (BBO), or the Decihertz Interferometer Gravitational Wave Observatory (DECIGO); all these observatories operate at some typical frequencies ranging from $10^{-4}$ Hz to $10^4$ Hz, and it is highly expected in the community that GW should be detected within this decade.

A number of constraints on GW have been derived from a variety of considerations related to things such as millisecond pulsars~\cite{pulsar} and Big Bang nucleosynthesis~\cite{BBN}. Indeed, there are hopes that cosmology may in near future offer more insight on the existence and nature of GW. From the observations of the Cosmic Microwave Background temperature and polarization~\cite{WMAP7} anisotropies, one can already infer upper bounds on the amplitudes of GW (see also~\cite{Elena}). 

During inflation metric perturbations, including tensor modes, i.e.~GW, are generated. If the scale of inflation is sufficiently high, the GW background from inflation could be detected directly~\cite{Kamionkowski} with satellite missions such as the Planck Surveyor or the proposed CMBpol satellite. Moreover, post-inflationary dynamics can also be a source of stochastic backgrounds of GW, generated by causal mechanisms very different from the quantum nature of the inflationary GW. After inflation the energy stored in the scalar inflaton field responsible for the superluminal expansion is converted into (almost) all the matter and radiation of the universe. This stage is called reheating, by the end of which the inflaton decay products have thermalized among themselves and the standard hot Big Bang evolution can commence. The process of reheating is not well understood. However, it is often assumed to be initially driven by non-perturbative effects, consisting in violent bursts of particle production known as preheating~\cite{preheating}. 

In this letter we will focus on the generation of GW right after inflation during preheating or other stages of non-perturbative particle production.
Most previous studies of post-inflationary phenomena generating GW have focused only on bosonic sources~\cite{TkachevGW,Dani,PhTs}, usually scalar fields. Here we want to complete the picture by considering fermionic fields as the source of GW. The phenomena in which fermions are created after inflation correspond to out-of-equilibrium periods in the evolution of the universe. Thus the created fermions have typically a non-thermal spectra and, as a consequence, they have a non-trivial anisotropic stress which will source GW. We will consider the possibility that the inflaton couples to fermions, and discuss the production of GW during fermionic preheating~\cite{Greene,Baacke}. Our considerations also apply to situations where fermions are produced by a scalar field other than the inflaton. The fermions we observe in Nature must have been generated sometime between the end of inflation and Big Bang nucleosynthesis. Fermionic preheating, or production of fermions in general, is thus no less natural than the usual bosonic production, but it is technically much more difficult to treat.

Since sources of GW produce spectra with distinctive shapes and amplitudes, it is important to characterize all the potential sources. The question then is: how big is the amplitude, and what is the frequency, of the stochastic GW produced by fermions in the early universe? The aim of the present letter is to answer these questions. We first develop a general formalism for computing the GW spectrum generated by an ensemble of fermions, and then apply it to two distinct scenarios: 1) fermions generated in preheating, and 2) thermal Universe into which the fermions are injected. From now on we will work in  units $\hbar = c = 1$, with $M_p \approx 2.4\times10^{18}\,{\rm GeV}$ the reduced Planck mass. Summation will be assumed over repeated indices.\\

\noindent{\bf Gravitational waves from fermions: the formalism.} Let us consider a flat Friedman-Robertson-Walker background with $a(t)$ the scale factor and $t$ conformal time. GW are the transverse-traceless (TT) part of metric perturbations $h_{ij}$,
\begin{equation}
ds^2 = a^2(t)(-dt^2+(\delta_{ij}+h_{ij})dx^idx^j),
\end{equation}
subject to $\partial_i h_{ij} = h_{ii} = 0$. The spectrum of energy density of a stochastic GW background in comoving momentum $k$ is given as
\begin{eqnarray}\label{eq:GWspectrum}
\frac{d\rho_{_{\rm GW}}}{d\log k}(k,t) &=& \frac{M_p^2k^3|{\dot h}_{k}(t)|^2}{8\pi^2a^2(t)}\,,
\end{eqnarray}
where $|{\dot h}_k(t)|^2$ is the power spectrum of $\dot h_{ij} = \frac{dh_{ij}}{dt}$.
For initial conditions $h_{ij}(t_i) = \dot h_{ij}(t_i) = 0$, 
the GW spectrum at sub-horizon scales becomes~\cite{Maggiore}
\begin{eqnarray}\label{eq:GWfromUETC}
\frac{d\rho_{GW}}{d\log k}(k,t) &=& \frac{k^3}{4\pi^2M_p^2}{1\over a^4(t)}\int_{t_i}^t dt' \int _{t_i}^t dt''a^3(t') a^3(t'') \nn\\
&& \times\,\, \cos(k(t'-t''))\,\Pi^2(k,t',t''),
\end{eqnarray}
where ${\Pi}^2$ is the unequal time correlator (UTC) of the TT-part of the anisotropic-stress $\Pi_{ij}^{\TT}$, 
\begin{equation}
\langle {\Pi}_{ij}^\TT\hspace*{-0.5mm}(\bk,t)\,{{\Pi}_{ij}^{\TT}}^{\hspace*{-0.2mm}*}\hspace*{-1mm}(\bk',t')\rangle = (2\pi)^3{\Pi}^2(k,t,t')\delta_D(\bk-\bk')
\end{equation}

Let us now consider spin-${1\over2}$ fermions as the source of GW, which can be represented as
\begin{equation}
\psi(\bx,t) = \hspace*{-1mm}\int\hspace*{-2mm}\frac{d\bk}{(2\pi)^3}e^{-i\bk\bx}\left\lbrace  a_{\bk,r}{\tt u}_{\bk,r}(t) + b^\dag_{-\bk,r}{\tt v}_{\bk,r}(t) \right\rbrace,
\vspace*{-0.5cm}
\end{equation}
$${\tt u}_{\bk,r} = (u_{\bk,+}S_r~~u_{\bk,-}S_r)^{\rm T}, {\tt v}_{\bk,r} = (v_{\bk,+}S_{-r}~~v_{\bk,-}S_{-r})^{\rm T},\nn
$$
with 
$r = 1,2$, $S_{1} = -S_{-2} = (1~~0)^{\rm T}$, $S_{2} = S_{-1} = (0~~1)^{\rm T}$, 
and $a_{r}, b_{r}$ the usual creation/annihilation operators obeying the relations $\lb a_{r}(\bk), a_{s}^\dag(\bq)\rb$ = $\lb b_{r}(\bk), b_{s}^\dag(\bq)\rb$ = $ (2\pi)^3\delta_{rs}\delta_D(\bk-\bq)$, $\lb a_{r}(\bk), b_{s}^\dag(\bq)\rb$ = 0. The fermion energy-momentum tensor is given by~\cite{BD}
\begin{eqnarray}
T_{ij}(\bx,t) = {1\over2a(t)}\left(\bar\psi\gamma_{(i}\overrightarrow{D}_{j)}\psi - \bar\psi\overleftarrow{D}_{(i}\gamma_{j)}\psi \right),
\end{eqnarray}
with $D_i \equiv \partial_i + {1\over4}[\gamma_\alpha,\gamma_\beta]\omega^{\ab}_j$ the covariant derivative, $\gamma_i$ the standard (flat-space) Dirac matrices, and $\omega^{\ab}$ the spin connection. The TT part of the anisotropic-stress in Fourier space, can be simply obtained by means of the orthogonal projector ${\mathcal P}_{ij} \equiv \delta_{ij} - \hat k_i \hat k_j$, as $\Pi_{ij}^\TT(\bk,t) = {\mathcal P}_{il}T_{lm}{\mathcal P}_{mj} - {1\over2}{\mathcal P}_{ij}{\mathcal P}_{lm}T_{lm}$. We then have all the ingredients to calculate $\Pi^2(k,t,t')$ and obtain the GW spectrum. Using the fact that $v_{\bp,\pm} = \pm u_{\bp,\mp}^*$, we find
\vspace*{-0mm}
\begin{eqnarray}\label{eq:GWspectrumII}
\frac{d\rho_\GW}{d\log k}(k,t) &=& {(k^3/M_p^2)\over 8\pi^4a^4(t)}\int \hspace*{-1mm}dp\,d\theta\,p^4{\rm sin}^3\theta\,F(k,p,\theta),\nn\\
F(k,p,\theta) &=& 
|I_{+}^{(c)}-I_{-}^{(c)}|^2 + |I_{+}^{(s)}-I_{-}^{(s)}|^2\\
I_{\pm}^{(c)}(k,p,\theta) &\equiv& \int\hspace*{-1mm}{dt'\over a(t')}\,{\rm cos}(kt')\, u_{\bk-\bp,\pm}(t')u_{\bp,\pm}(t'),\nn
\end{eqnarray}
with $I^{(s)}_{\pm}$ defined analogously by replacing $\cos(kt)$ by $\sin(kt)$. Eq.~(\ref{eq:GWspectrumII}) is the master set of formulae that describe the GW spectrum at subhorizon scales, as generated by some fermionic field $\psi$ with eigenfunctions $u_{\bk,\pm}(t)$. The $eom$ for $u_{\bk,\pm}(t)$ will follow from the Dirac equation. For any  process in the early Universe where fermions are excited, one just needs to plug in the solutions $u_{\bk,\pm}(t)$ into the master equation (\ref{eq:GWspectrumII}) to find the spectrum of GW.

The structure of the formulae in Eq.~(\ref{eq:GWspectrumII}) resembles that of scalar fields sourcing GW. However, in the bosonic case, apart from multiplicative factors, there appears a power $p^6$  instead of $p^4$ in the integrand, there are no polarization indices $+,-$, and of course the fermionic mode functions $u_{\bk,\pm}(t)$ are replaced by the Klein-Gordon scalar modes $\phi_{\bk}(t)$.

Both bosonic and fermionic vacuum expectation values (VEVs), like $\Pi^2$, require regularization. 
In the case of bosons, this has not been an issue in the literature, since the bosonic UTC are either introduced as a theoretical regularized Ansatz, or in the case of lattice simulations, the ultraviolet (UV) modes causing the divergence are simply not captured. In the fermionic case one cannot skip regularization. To regularize $\Pi^2$, note first that the VEV of the source itself, $\Pi_{ij}^\TT$, needs also to be regularized. 
Similarly to the flat-space case, regularization of the source's VEV amounts to a substraction of the zero-point fluctuations or, equivalently, to a time-dependent normal-ordering ($tNO$) procedure, i.e.~$\langle \Pi_{ij}^{TT} \rangle_{\rm reg} \equiv \langle 0| \Pi_{ij}^{TT} | 0 \rangle - \langle 0_t| \Pi_{ij}^{TT} | 0_t \rangle$, with $|0\rangle$ the initial vacuum and $|0_t\rangle$ the vacuum at time $t$. This removes the unphysical divergence in the VEV of $\Pi_{ij}^{TT}$ at every time $t$. In practice, $tNO$ amounts to the replacement $(u_{\bp,\pm}u_{\bp',\pm})\to$ 
$(u_{\bp,\pm}u_{\bp',\pm})_{_{\rm reg}} = |\beta_\bp||\beta_{\bp'}|u_{\bp,\pm}u_{\bp',\pm}-(\beta_\bp\beta_{\bp'}u_{\bp,\mp}u_{\bp',\mp})^*$
with $|\beta_\bp| = \sqrt{1-|\alpha_\bp|^2}$, where $\alpha$, $\beta$ are the canonical Bogoliubov coefficients~\cite{BD} connecting the initial creation and annhilation operators with those at time $t$. To render $\Pi^2$ finite we just need to replace in Eq.~(\ref{eq:GWspectrumII}) the functions $I^{(c)}_{\pm}$ by
\begin{eqnarray}
{\mathcal I}^{(c)}_{\pm} \equiv \int\hspace*{-1mm}{dt'\over a(t')}\,{\rm cos}(kt')\, \left(u_{\bk-\bp,\pm}(t')u_{\bp,\pm}^*(t')\right)_{\rm reg},
\end{eqnarray}
and similarly for ${\mathcal I}^{(s)}_{\pm}$. 
In this way the convergence of the integration over $p$ is ensured by the suppression of the UV divergent modes.
\\

\noindent{\bf Stochastic background of gravitational waves from fermions produced in the early Universe.} Several scenarios of the early Universe may create high-energy out-of-equilibrium fermions by non-perturbative effects, for instance a homogeneous scalar field oscillating around the minimum of a potential. 
In such situation, if fermions are coupled to that field, a non-perturbative population of their modes (respecting Pauli-blocking) takes place~\cite{Greene,Baacke}. This generates a non-trivial anisotropic-stress that sources a stochastic background of GW. In this letter we consider two different scenarios:

I) {\it (p)reheating after inflation}: Let us assume an inflaton oscillating around the minimum of its potential after the end of inflation. This is the case of chaotic inflation models with polynomial potentials like $V \propto \phi^2$ or $V \propto \phi^4$. In general, the shape of the inflaton potential during inflation is however irrelevant for our purposes. We will just assume that inflation took place at some energy scale $E_I$, and that  afterwards the inflaton 
oscillates with a certain frequency 
and decreasing amplitude. 
The expansion of the Universe will be dictated during (p)Reheating by the inflaton energy density. For a polynomial potential $V(\phi)$ it is well known that the scale factor behaves as a power law in time. In the simplest case $V(\phi) = {1\over2}\omega_0^2\phi^2$, $a(t) \propto t^2$, so the expansion of the Universe is matter-dominated (MD).

II) {\it Oscillating scalar field in a thermal era}: It is also possible to consider a scalar field, other than the inflaton, which oscillates coherently around a potential minimum when the Universe has already entered into a thermal era. That is the case, for example, of the curvaton scenario~\cite{curvatons}, which is an alternative to single-field inflationary models. After the Universe has reheated, the curvaton 
oscillates with decreasing amplitude and
fixed frequency. The expansion of the Universe is radiation-dominated (RD), driven by the thermal relativistic bath of particles with energy density $\rho_{\rm th} \propto 1/a^4$, and $a(t) \propto t$. We assume oscillations begin at some energy scale $E_I$ way above the electroweak scale.

In both scenarios the oscillatory field $\phi$ behaves as a damped oscillator, $\phi(t)=\Phi(t)F(t)$, with decreasing amplitude $\Phi(t)$ and periodic behaviour $F(t+2\pi/\omega_0) = F(t) \leq 1$, as determined by the choosen potential. We will assume that $\phi$ is coupled to some fermion species $\psi$ via a Yukawa interaction $h\phi\bar\psi\psi$, with $h$ the interaction strength. The Dirac equation yields the equation for the mode functions $u_{\bk,\pm}$ as
\begin{eqnarray}\label{diracmodes}
\ddot{u}_{\bk,\pm} + \left(\bk^2+h^2a^2\phi^2 \pm ih(\dot a\phi + a\dot\phi)\right)u_{\bk,\pm} = 0~.
\end{eqnarray}
We have solved numerically the mode equations (\ref{diracmodes}) for the scenarios I and II, scanning for the parameters $h$, $\omega_0$ and $E_I$ and assuming $V(\phi) = {1\over2}\omega_0^2\phi^2$ and initial conditions corresponding to a vanishing particle density. In both scenarios, each time $\phi$ passes through the minimum of the potential, fermions are created out-of-equilibrium by non-perturbative parametric effects~\cite{preheating}.

Pauli blocking prevents the fermion occupation numbers to grow arbitrarily. Fermion excitations are forced to fill up a ``Fermi-sphere'' in momentum space of comoving radius $k_F \sim a(t)q^{1/4}\omega_0$, with $q \equiv h^2(\Phi/\omega_0)^2$ the resonant parameter~\cite{Greene}. The occupation numbers with momenta smaller than $k_F$ oscillates continuously between 0 and 1, whilst the radius $k_F$ grows as $\propto a^{1/4}$. When the energy density of the created fermions has grown to a significant fraction of the energy of the oscillating field $\phi$ -- usually within tens of oscillations --, the creation of fermions finally ceases. The excitations of the fermionic modes within the Fermi-sphere source the generation of GW. The non-excited modes, i.e.~those outside the sphere, are on the contrary responsible for the UV divergence discussed before. The regularization scheme we impose filters the infrared (IR) modes within the sphere $k \lesssim k_F$ and removes the contribution from the UV ones. 

It is useful to define the times $t_I$, $t_*$ and $t_{\rm RD}$, as the initial time, the end of GW production, and the first moment when the Universe becomes RD, respectively. In the scenario I, $t_I < t_{*} < t_{\rm RD}$, so in this case the effective equation of state $p/\rho = w$ between $t_I$ and $t_{\rm RD}$ is typically different from that of RD, $w_{\rm RD} = 1/3$ (unless $V(\phi) \propto \phi^4$). It is then convenient to introduce the factor $\epsilon \equiv (a_*/a_{\rm RD})^{(1-3w)}$, see next. In the scenario II, $t_{\rm RD} \leq t_I < t_*$, so that $\epsilon = 1$. Below the Planck scale, GW decouple immediately after production, so we can evaluate the GW energy density spectrum today from the spectrum computed at the time of production. We just have to redshift the amplitude and wavenumbers,
\begin{eqnarray}\label{eq:freqAndAmplitude}
f &=& \epsilon^{1/4}\left({a_I\over a_*}\right)\left(\omega_0\over \rho_*^{1/4}\right)\left({k\over\omega_0}\right)\times5\cdot10^{10}~{\rm Hz}\nn\\
h^2\Omega_\GW &\equiv& {h^2\over\rho_c}\hspace*{-1mm}\left({d\rho_\GW\over d\log k}\right)_{\hspace*{-1mm}0} = h^2\Omega_{\rm rad}\left(g_0\over g_*\right)^{\hspace*{-1mm}{1\over3}}{\epsilon\over \rho_*}\hspace*{-1mm}\left({d\rho_\GW\over d\log k}\right)_{\hspace*{-1mm}*},\nn
\end{eqnarray}
with $h^2\Omega_{\rm rad} \approx 4\times10^{-5}$, $\rho_* = E_I^4(a_I/a_*)^{3(1+\omega)}$ the energy density at $t_*$, and $(g_0/g_*)^{1/3} \sim \mathcal{O}(0.1)$ the ratio of the number of relativistic $dof$ today to those active at $t_*$.

\begin{figure}[t]
\begin{center}
\includegraphics[width=8cm,height=5cm,angle=0]{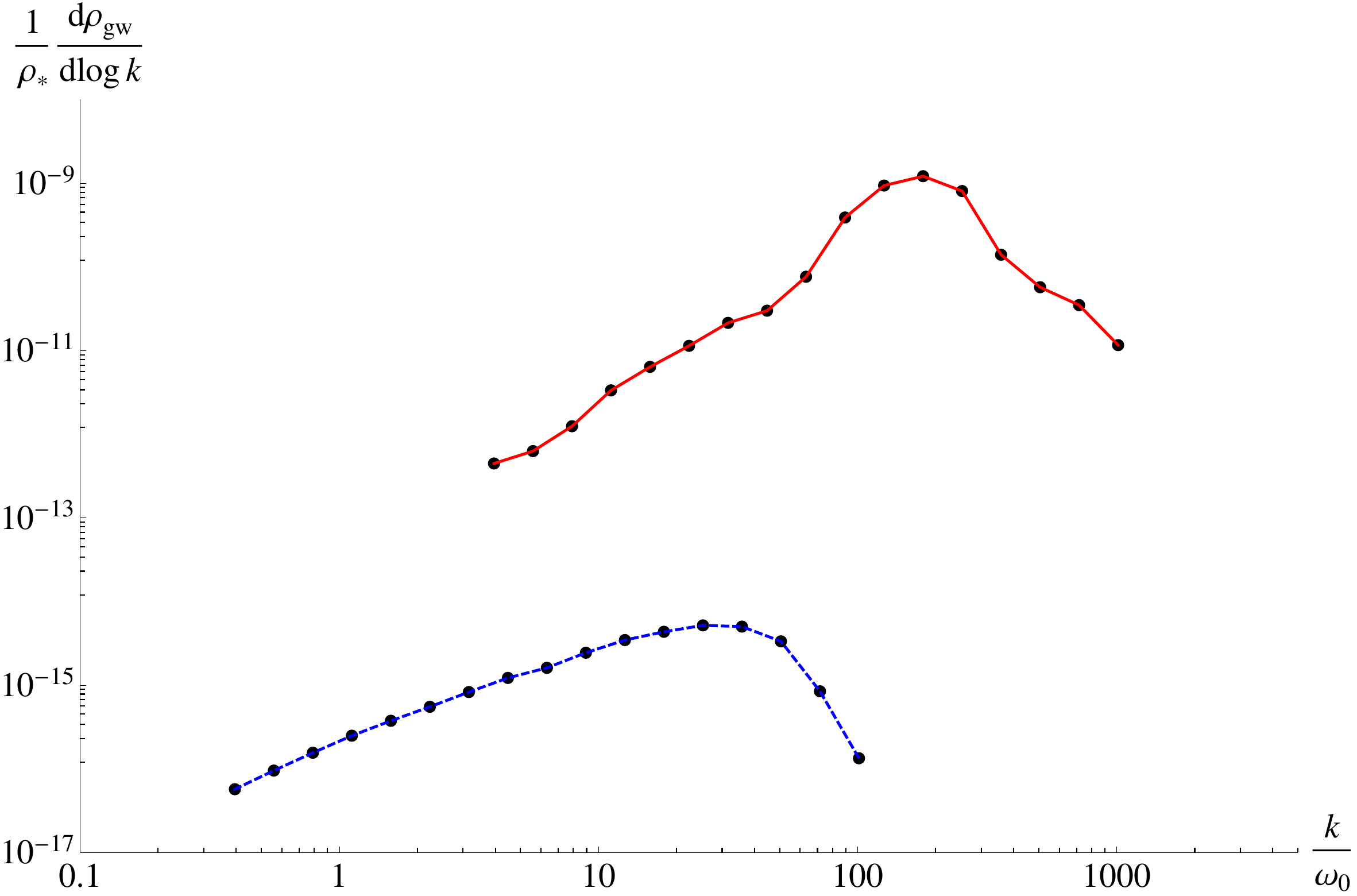}
\end{center}
\vspace*{-5mm}
\caption{Spectra of GW at the time of production, in the scenario I with $E_I \sim 10^{16}$ GeV and $\omega_0 \sim 10^{14}$ GeV, for $q = 10^2$ (dashed line, peaked at $k = 8q^{1/4}\omega_0$) and $q = 10^{6}$ (solid line, peaked at $k = 6q^{1/4}\omega_0$). The spectra in the scenario II for the same parameters look qualitatively the same.}
\label{fig1}
\end{figure}

In Fig.~\ref{fig1} we show two examples of GW energy density spectra, obtained with the machinery presented in the previous section. The spectra grows in the IR region, reaches a maximum at $k \sim q^{1/4}\omega_0$, and finally falls down in the UV region. The spectra are peaked, as expected on dimensional grounds, around the scale $\sim k_F$, characteristic of the fermions' dynamics. From the $k\rightarrow0$ limit in Eq.~(\ref{eq:GWspectrumII}), we would expect the spectra to grow as $\propto k^3$ in the very IR region. Instead we observe a behaviour $\sim k^2$, probably signaling that we are exploring $k$'s too close to the maximum, and that one should look for much smaller $k$'s to probe the $k^3$ behaviour. Unfortunately, due to computer limitations, both the limits $k \ll k_F$ and $k \gg k_F$ are indeed very challenging to probe, so we leave a complete characterization of the IR and UV tails of the GW spectra for a future publication.

With our formalism we can obtain nevertheless the most important aspects characterizing any GW spectrum: the spectral shape around the peak, and the amplitude and position of such peak. For instance in the scenario I with $E_I \sim 10^{16}$ GeV and $\omega_0 \sim 10^{14}$ GeV, we find for $q = 10^2$, today's peak frequency and amplitude as $f \sim 10^{9}$ Hz, $h^2\Omega_{GW} \sim 10^{-20}$, and for $q = 10^6$, as $f \sim 10^{10}$ Hz, $h^2\Omega_{GW} \sim 10^{-14}$. In the scenario II, for the same $E_I$ and $\omega_0$, we find $f \sim 10^9$ Hz, $h^2\Omega_{GW} \sim 10^{-18}$ for $q = 10^2$, and $f \sim 10^{10}$, $h^2\Omega_{GW} \sim 10^{-12}$ for $q = 10^6$. Thus the amplitude of the GW background explored here can indeed be very significant. However the typical frequencies are too large as compared to the range probed by GW observatories ($\sim 10^{-4}-10^4$ Hz). From Eq.~(\ref{eq:freqAndAmplitude}) we learn that if $\omega_0/\rho_I^{1/4} \ll 1$, we could decrease significantly the frequency towards the observable window. However $\Omega_{\rm GW}$ is suppressed by $(\omega_0^4/\rho_I)$, so the amplitude would be far too small, $\Omega_{GW} \ll 10^{-20}$. We hope to return to a systematic exploration of the parameters in a future publication.

There are some indications for scenarios where a GW background at sufficiently low frequencies and high enough amplitudes could be found. For instance, in hybrid inflation, the frequency $\omega_0$ should be replaced by $\sqrt{\lambda}v$, and the initial energy scale $E_I$ by $\sim \lambda^{1/4}v$, where $\lambda$ and $v$ are the self-coupling and VEV of an auxiliary field coupled to the inflaton. The frequencies would then scale as $f \propto \lambda^{1/4}$~\cite{Dani} so that for sufficiently small $\lambda$, one could obtain a peak frequency within the observable range. This possibility can be explored by using the machinery developed in this Letter.

Summarizing, we have shown that fermions in the early Universe may be very efficient generators of GW. These waves remain decoupled since the moment of their production, and thus the amplitude and shape of their spectrum probes the physics responsible for their generation. The characteristic spectrum is different from other backgrounds of GW, like those arising from binaries coalescing~\cite{Maggiore}, which are decreasing with frequency, or those arising from inflation or self-ordering fields~\cite{GWfromSOSF}, which are flat. A comparison of the exact shape of the spectra predicted here versus those generated by scalar fields in phase transitions and (p)reheating requires further study. Here we just want to emphasize that, contrary to naive expectations based on the Pauli principle, we found that fermions are capable of producing a stochastic background of GW with a very large amplitude. Although it may be manifest only at very high frequencies much beyond present/planned GW detectors' sensitivities, there is however some hope that small-coupling models could give rise to a signal in observable frequency ranges.

{\it Acknowledgments} We thank R. Durrer, J. Garc\'ia-Bellido and A. Riotto for useful comments. TM is supported by the Magnus Ehrnrooth foundation, while KE and DGF are respectively supported by the Academy of Finland grants 131454 and 218322.

\end{document}